%% file: TDSL2.tex
\documentclass[sigplan,10pt]{acmart}
\usepackage{multicol}
\renewcommand\footnotetextcopyrightpermission[1]{} 
\pdfoutput=1

\acmConference[Arxiv]{Arxiv}{2020}{}
\acmYear{2020}
\acmISBN{} 
\acmDOI{} 
\startPage{1}
\setcopyright{none}
\startPage{1}

\setcopyright{none}

\usepackage{booktabs}   
\usepackage{subcaption} 

\usepackage[utf8]{inputenc}
\usepackage{microtype}

\usepackage{comment}

\usepackage{url}
\usepackage{hyperref}

\usepackage{color}
\definecolor{pblue}{rgb}{0.13,0.13,1}
\definecolor{pgreen}{rgb}{0,0.5,0}
\definecolor{pred}{rgb}{0.9,0,0}
\definecolor{pgrey}{rgb}{0.46,0.45,0.48}
\usepackage{listings}
\newcommand{\var}[1]{\mathit{#1}}
\newcommand{\estm}{$\varepsilon$-STM}
\newcommand{\vcc}{PSTM}

\usepackage{algorithm}
\usepackage[noend]{algpseudocode}

\algnewcommand\algorithmicswitch{\textbf{switch}}
\algnewcommand\algorithmiccase{\textbf{case}}
\algnewcommand\algorithmicassert{\texttt{assert}}
\algnewcommand\Assert[1]{\State \algorithmicassert(#1)}%
\algdef{SE}[SWITCH]{Switch}{EndSwitch}[1]{\algorithmicswitch\ #1\ \algorithmicdo}{\algorithmicend\ \algorithmicswitch}%
\algdef{SE}[CASE]{Case}{EndCase}[1]{\algorithmiccase\ #1}{\algorithmicend\ \algorithmiccase}%
\algtext*{EndSwitch}%
\algtext*{EndCase}%

\algdef{SE}[SUBALG]{Indent}{EndIndent}{}{\algorithmicend\ }%
\algtext*{Indent}
\algtext*{EndIndent}
\newcommand{\remove}[1]{}

\begin{document}

\title[Nesting in Transactional
Libraries]{Using Nesting to Push the Limits of\\ Transactional Data
Structure Libraries}
\thanks{This work was partially supported by an Israel Innovation Authority Magnet Consortium (GenPro).}


\author{Gal Assa}
\affiliation{
 \institution{Technion}
}
\email{galassa@campus.technion.ac.il}
\author{Hagar Meir}
\affiliation{
\institution{IBM Research} 
}
\email{hagar.meir@ibm.com}
\author{Guy Golan-Gueta}
\affiliation{
\institution{Independent researcher} 
}
\email{guyg.guyg@gmail.com}
\author{Idit Keidar}
\affiliation{
\institution{Technion} 
}
\email{idish@ee.technion.ac.il}
\author{Alexander Spiegelman}
\affiliation{
\institution{Independent researcher} 
}
\email{sasha.spiegelman@gmail.com}

\begin{abstract}

Transactional data structure libraries (TDSL) combine the ease-of-programming of transactions with the high performance and scalability of custom-tailored concurrent data structures.
They can be very efficient thanks to their ability to exploit data structure semantics in order to reduce overhead, aborts, and wasted work compared to general-purpose software transactional memory.
However, TDSLs were not previously used for complex use-cases involving long transactions and a variety of data structures.

In this paper, we boost the performance and usability of a TDSL, towards allowing it to support complex applications.
A key idea is \textit{nesting}.
Nested transactions create checkpoints within a longer transaction,
so as to limit the scope of abort, without changing the semantics of the original transaction.
We build a Java TDSL with built-in support for nested transactions over a number of data structures.
We conduct a case study of a complex network intrusion
detection system that invests a significant amount of
work to process each packet. Our study shows that our library
outperforms publicly available STMs twofold without nesting, and by up to 16x when
nesting is used.

%
%
\end{abstract}



\maketitle

\input{intro}
\input{bg}
\input{nestingInTDSL}
\input{NIDS}

\input{additionalDS}

\input{Evaluation}

\input{related_work}
\input{Conclusion}


\clearpage
\bibliography{references}
\clearpage
\input{DISCAppendix}

\end{document}

%% file: intro.tex
\section{Introduction}

\subsection{Transactional Libraries}

The concept of memory
transactions~\cite{herlihy1993transactional} is broadly considered to
be a programmer-friendly paradigm for writing concurrent
code~\cite{herlihy2005transactional, Scott:Transactional}.
A transaction spans multiple operations, which appear to
execute atomically and in isolation, meaning that either all
operations commit and affect the shared state or the transaction
aborts.
Either way, no partial effects of on-going transactions
are observed.

Despite their appealing ease-of-programming, software transactional
memory (STM) toolkits~\cite{TL2,
herlihy2003software, perelman2011smv} are
seldom deployed in real systems due to their huge performance
overhead. 
The source of this overhead is twofold. 
First, an STM needs to monitor all random memory accesses made in the course of
a transaction (e.g., via instrumentation in VM-based languages~\cite{deuceTL2}), and second, STMs abort transactions due to conflicts.
Instead, programmers widely use concurrent data structure
libraries~\cite{shavit2000skiplist, lea2000concurrent, heller2005lazy, bronson2010practical}, which are much faster but guarantee atomicity only at the level of a single operation on a single data structure.

To mitigate this tradeoff,
Spiegelman et al.~\cite{TDSL} have proposed \emph{transactional data
structure libraries (TDSL)}.
In a nutshell, the idea is to trade generality for performance.
A TDSL restricts transactional access to a pre-defined set of
data structures rather than arbitrary memory locations, which eliminates the need for instrumentation. Thus, a TDSL can exploit the data structures' semantics and structure to get
efficient transactions bundling a sequence
of data structure operations.
It may further manage aborts on a semantic level, e.g.,
two concurrent transactions can simultaneously change two different
locations in the same list without aborting.
While the original TDSL library~\cite{TDSL} was written in
C++, we implement our version in Java. 
We offer more background on  TDSL in Section~\ref{bg}.

Quite a few  works ~\cite{FL-wait-free1, loft,FL-lock-free1, defer} have used and
extended TDSL and similar approaches like STO ~\cite{itiswhatitis} and transactional boosting \cite{herlihy2008transactional}.
These efforts have shown good performance for fairly short transactions
on a small number of data structures.
Yet, despite their improved scalability compared to general
purpose STMs, TDSLs have also not been applied to long transactions or
complex use-cases.
A key challenge arising in long transactions is the high potential
for aborts along with the large penalty that such aborts induce as
much work is wasted.

\subsection{Our Contribution}

\textbf{Transactional nesting.} In this paper we push the limits of the TDSL concept in an attempt to make it more broadly applicable. Our main contribution, presented in Section~\ref{nTDSL}, is facilitating long transactions via \emph{nesting}~\cite{moss1981nested}.
Nesting allows the programmer to define nested \emph{child}
transactions as self-contained parts of larger \emph{parent}
transactions. This controls the program flow by
creating \textit{checkpoints}; upon abort of a nested child
transaction, the checkpoint enables retrying only the child's part and not the
preceding code of the parent. This reduces wasted work, which, in
turn, improves performance. At the same time, nesting does not relax consistency or isolation, and continues to ensure that the entire parent transaction is executed atomically. We focus on \emph{closed nesting}~\cite{turcu2012closed}, which, in contrast to so-called flat nesting, limits the scope of aborts, and unlike open nesting~\cite{MOSSHOSKING}, is generic and does not require semantic
constructs.

The flow of nesting is shown in Algorithm
\ref{alg:flow}. When a child commits, its local state is
migrated to the parent but is not yet reflected in
shared memory. If the child aborts, then the parent transaction is
checked for conflicts.
And if the parent incurs no conflicts in its part of the code, then
only the child transaction retries. Otherwise, the entire transaction
does.
It is important to note that the semantics provided by the parent
transaction are not altered by nesting.
Rather, nesting allows programmers to identify parts of the code that
are more likely to cause aborts and encapsulate them in child
transactions in order to reduce the abort rate of the parent.

\begin{algorithm}[tb]
\caption{Transaction flow with nesting}
\label{alg:flow}
\begin{algorithmic}[1]\small

\State TXbegin()
	\Indent
   \State [Parent code]\Comment{On abort -- retry parent} 
   \State nTXbegin() \Comment{Begin child transaction}
   		\Indent
    	\State [Child code] \Comment{On abort -- retry child or parent}
		\EndIndent
   \State nTXend() \Comment On commit -- migrate changes to parent
   \State [Parent code]\Comment{On abort -- retry parent} 
   \EndIndent

\State TXend() \Comment{On commit -- apply changes to thread state} 

\end{algorithmic}
\end{algorithm}



Yet nesting induces an overhead which is not always offset by its benefits. We investigate this tradeoff using microbenchmarks. We find that nesting is helpful for highly contended operations that are likely to succeed if retried. Moreover, we find that nested variants of TDSL improve performance of state-of-the-art STMs with transactional friendly data structures. 

\textbf{NIDS benchmark.} In Section~\ref{NIDS} we introduce a new benchmark of a \emph{network intrusion detection system (NIDS)}~\cite{Intruder}, which invests a fair amount of work to process each packet.
This benchmark features a pipelined architecture with long transactions, a variety of data structures, and multiple points of contention. It follows one of the designs suggested in \cite{Intruder} and executes significant computational operations within transactions, making it more realistic than existing intrusion-detection benchmarks (e.g., \cite{JSTAMP,STAMP}). 

\textbf{Enriching the library.} In order to support complex applications like NIDS, and
more generally, to increase the usability of TDSLs, we
enrich our transactional library in Section~\ref{MoreDS} with additional data structures -- producer-consumer pool, log, and stack -- all of which support nesting.
The TDSL framework allows us to custom-tailor to each data structure its own concurrency control mechanism.
We mix optimism and pessimism (e.g., stack operations
are optimistic as long as a child has popped no more than it pushed,
and then they become pessimistic), and also fine tune the granularity
of locks (e.g., one lock for the whole stack versus one per slot in the producer-consumer pool).

%

\textbf{Evaluation.} In Section \ref{Evaluation}, we evaluate our NIDS application. 
We find that nesting can improve performance by up to 8x.
Moreover, nesting improves scalability, reaching peak
performance with as many as 40 threads as opposed to 28 without
nesting.

%

\textbf{Summary of contributions.} 
This paper is the first to bring nesting into transactional data
structure libraries and also the first to implement closed nesting in sequential STMs.
We implement a Java version of TDSL with built-in support for
nesting. 
Via microbenchmarks, we explore when nesting is beneficial and show that in some scenarios, it can
greatly reduce abort rates and improve performance.
We build a complex network intrusion detection application, while enriching our library with the data structures required to support it.
We show that nesting yields significant improvements in performance and abort rates. 


%% file: bg.tex
\section{A Walk down Transactional Data Structure Lane}\label{bg}

Our algorithm builds on ideas used in TL2~\cite{TL2}, which is a generic STM framework, and in
TDSL~\cite{TDSL}, which suggests forgoing generality for increased efficiency. We briefly overview their modus operandi as
background for our work. 
\par
The TL2~\cite{TL2} algorithm introduced a
version-based approach to STM. The algorithm's building
blocks are version clocks, read-sets, write-sets, and a per-object
lock. A \emph{global version clock (GVC)} is shared among all
threads. A transaction has its own \emph{version clock (VC)}, which
is the value of GVC when the transaction begins. A shared object has
a version, which is the VC of the transaction that most recently
modified it. The read- and write-sets consist of references to
objects that were read and written, respectively, in a transaction's
execution. 

Version clocks are used for \textit{validation}: Upon read,
the algorithm first checks if the object is locked and then the VC of the read
object is compared to the transaction's VC. If the object is locked
or its VC is larger than the transaction's, then we say the validation \emph{fails},
and the transaction
aborts. Intuitively, this indicates that there is a \emph{conflict}
between the current transaction, which is reading the object, and a
concurrent transaction that writes to it. 

At the end of a transaction, all the objects in its write-set are locked and then every object in the read-set is revalidated. If this succeeds, the transaction commits and its write-set is reflected to shared memory. If any lock cannot be obtained or any of the objects in the read-set does not pass validation, then the transaction aborts and retries. 

\emph{Opacity}~\cite{OPACITY} is a safety property that
requires every transaction (including aborted ones) to observe only \emph{consistent} states of the system that could have been observed in a sequential execution. TL2's read-time validation (described above) ensures opacity. 

In TDSL, the TL2 approach was tailored to specific data structures (skiplists and queues) so as to benefit from their
internal organization and semantics. TDSL's skiplists use small read- and write-sets capturing only accesses that 
induce conflicts at the data strucutre's semantic level. For example, whereas TL2's read-set holds all nodes traversed during the lookup of a particular key, TDSL's read-set keeps only the node holding this key. 
In addition, whereas TL2 uses only optimistic concurrency-control (with commit-time locking), TDSL's queue uses a semi-pessimistic approach. Since the head of a queue is a point of contention, \textit{deq} immediately locks the shared queue (although the actual removal of the object from the queue is deferred to commit time); the \textit{enq} operation remains optimistic.

Note that TDSL is less general than generic STMs: STM transactions span all memory accesses within a transaction, which is enabled, e.g., by instrumentation of binary code~\cite{angel2001byte} and results in large read- and write-sets. TDSL provides transactional semantics within the confines of the library's data structures while other memory locations are not accessed transactionally. This eliminates the need for instrumenting code. 

%


%% file: nestingInTDSL.tex
\section{Adding Nesting to TDSL}\label{nTDSL}

We introduce nesting into TDSL. Section \ref{ssec:nesting-scheme} describes the correct behavior of nesting and offers a general scheme for making a transactional \emph{data structure (DS)} nestable. Section \ref{ssec:nested-q-list} then demonstrates this technique in the two DSs supported by the original TDSL -- queue and skiplist. We restrict our attention to a single level of nesting for clarity, as we
could not find any example where deeper nesting is useful, though deeper nesting could be supported along the same lines if required. In Section \ref{ssec:ubenchmark} we use microbenchmarks to investigate when nesting is useful and when less so, and to compare our library's performance with transactional data structures used on top of general purpose STMs.

\subsection{Nesting Semantics and General Scheme}
\label{ssec:nesting-scheme}

Nesting is a technique for defining  child sub-transactions within a  transaction. 
A child 
has its own local state (read- and write-sets), and it may also observe its parent's local state.
A  child transaction's commit migrates its local state to its parent but not to shared memory visible by other threads. Thus, the child's operations take effect when the parent commits, and until then remain unobservable. 

\textbf{Correctness.} A nested transaction implementation ought to ensure that (1) nested operations are not visible in the shared state until the parent commits; and (2) upon a child's commit, its operations are correctly reflected in the parent's state exactly as if all these operations would have been executed as part of the parent. In other words, nesting part of a transaction does not change its externally visible behavior. 

\textbf{Implementation scheme.} In our approach, the child  uses its parent's VC. This way, the child and the parent observe the shared state at the same ``logical time'' and so read validations 
ensure that the combined state observed by both of them is consistent, as required for opacity.

Algorithm~\ref{alg:nestOps} introduces general primitives for nesting arbitrary DSs. The \textit{nTXbegin} and \textit{nCommit} primitives are exposed by the library and may be called by the user as in Algorithm \ref{alg:flow}. When user code operates on a transactional DS managed by the library for the first time, it is registered in the transaction's childObjectList, and its local state and lockSet are initialized empty.
\textit{nTryLock} may be called from within the library, e.g., a nested dequeue calls \textit{nTryLock}. Finally, \textit{nAbort} may be called by both the user and the library.

We offer the \textit{nTryLock} function to facilitate pessimistic concurrency control (as in TDSL's queues), where a lock is acquired before the object is accessed. This function (1) locks the object if it is not yet locked; and (2) distinguishes newly acquired locks from ones that were acquired by the parent. The latter allows the child to release its locks without releasing ones acquired by its parent.

A nested commit, \textit{nCommit}, validates the child's read-set in all the transaction's DSs \emph{without} locking the write-set. If validation is successful, the child migrates its local state to the parent, again, in all DSs, and also makes 
its parent the owner of all the locks it holds. To this end, every nestable DS must support \emph{migrate} and \emph{validate} functions, in addition to nested versions of all its methods.

\begin{algorithm}[htb]
\begin{algorithmic}[1]\footnotesize
\Procedure{nTXbegin}{}

\State alloc childObjectList, init empty
\EndProcedure

\State \textbf{On first access to obj in child transaction}
\Indent  \Indent
	add obj to childObjectList
\EndIndent  \EndIndent

\Procedure{nCommit}{}
\ForAll{obj in childObjectList}
	\State validate obj with parent's VC
	\If{validation fails}
		\State \textbf{nAbort}
	\EndIf
\EndFor
	\ForAll{obj in childObjectList}
		\State obj.migrate \Comment{DS specific code}
	\EndFor
	\ForAll{lock in lockSet}
		\State transfer lock ownership to parent
	\EndFor		
\EndProcedure

\Procedure{nAbort}{}
	\ForAll{obj in childObjectList}
		\State release locks in lockSet
	\EndFor
	\State parent VC $\gets$ GVC \label{line:vc-update}
	\ForAll{obj in childObjectList}
		\State validate parent \label{inalg:validateParent} \Comment{DS specific code}
		\If{}validation fails
			\State \textbf{abort} \Comment Retry parent
		\EndIf \label{line:end-abort}
	\EndFor
		 \State restart child\label{inalg:retryChild}
\EndProcedure

\Procedure{nTryLock}{obj}
\State \textbf{atomic} \Comment{Using CAS for atomicity}
\Indent
	\If{obj is unlocked}
		\State \textbf{lock} obj with child id
		\State \textbf{add} obj to lockSet
	\EndIf	
		\If{obj is locked but not by parent}
			\State \textbf{nAbort} \Comment Abort child
	
	\EndIf
\EndIndent
\EndProcedure

\end{algorithmic}
\caption{Nested begin, lock, commit, and abort}
\label{alg:nestOps}
\end{algorithm}

In case the child aborts, it releases all of its locks. Then, we 
need to decide whether to retry the child or abort the parent too.
Simply retrying the child without changing the VC is liable to fail because it would re-check 
the same condition during validation, namely, comparing read object VCs to the transaction's VC.  
We therefore update the VC to the current GVC value 
(line~\ref{line:vc-update}) before retrying. 
This ensures that the child will not re-encounter past conflicts. 
But in order to preserve opacity, we must verify that the state the parent observed is still consistent at 
the \emph{new} logical time (in which the child will be retried) because operations within a child transaction ought to be seen as if they were executed as part of the parent. 
To this end, we  revalidate the parent's read-set against the new VC (line \ref{inalg:validateParent}). This is done without locking its write-set. Note that if this validation fails then the parent is deemed to abort in any case, and the early abort improves performance. If the revalidation is successful, we restart only the child (line \ref{inalg:retryChild}).

Recall that retrying the child is only done for performance reasons and it is always safe to abort the parent. Specific implementations may thus choose to limit the number of times a child is retried.

\subsection{Queue and Skiplist}
\label{ssec:nested-q-list}

We extend TDSL's DSs with nested transactional operations in Algorithm \ref{alg:cQ}. 

The original queue's local state includes a list of nodes to enqueue and a reference to the last node to have been dequeued (together, they replace the read- and write-sets). We refer to these components as the parent's \textit{local queue}, or \textit{parent queue} for short. 
Nested transactions hold an additional \textit{child queue} in the same format.

\begin{algorithm}[h]
\begin{algorithmic}[1]\footnotesize
\State \textbf{Queue}
\Indent 
 \State sharedQ \Comment{Shared among all threads}
 \State parentQ, childQ \Comment{Thread local}
\EndIndent

\Procedure {nEnq}{val}
     \State childQ.\Call{append}{val} \label{inalg:cEnq}
\EndProcedure
\Procedure {migrate}{}
	\State \Call{parentQ.appendAll}{childQ}
\EndProcedure
\Procedure {validate}{}
	\State \Return true
\EndProcedure
\Procedure {nDeq}{\null} \label{inalg:deq}
	\State \Call{nTryLock}{\null} \label{inalg:deqb}
	\State val $\gets$ next node in sharedQ \Comment{stays in sharedQ} \label{inalg:deqS}
	\If {val = $\bot$}
		\State val $\gets$ next node in parentQ \Comment{stays in parentQ} \label{inalg:deqP}
		\If {val = $\bot$}
			\State val $\gets$ childQ.\Call{deq}{\null} \Comment{Removed from childQ} \label{inalg:deqe}
		\EndIf
	\EndIf
	\Return val
\EndProcedure


\end{algorithmic}
\caption{Nested operations on queues} 
\label{alg:cQ}
\end{algorithm}

\begin{figure}[h]
\centering
\includegraphics[width=0.8\columnwidth]{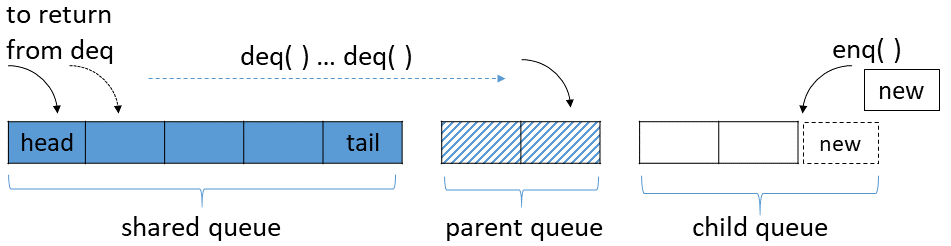}
\centering 
\includegraphics[width=0.8\columnwidth]{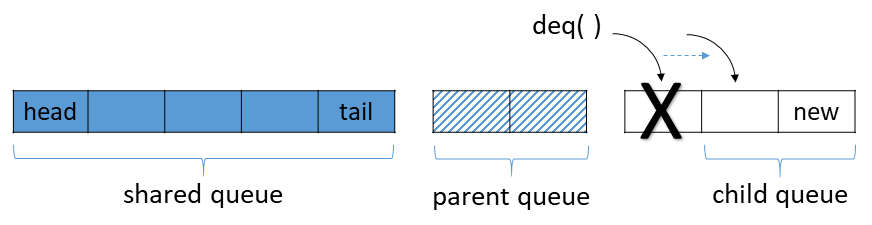}
\caption{Nested queue operations: deq returns objects from the shared, and then parent states without dequeuing them, and when they are exhausted, dequeues from the child's queue; enq always enqueues to the child's queue.}
\label{fig:nested-queue}
\end{figure}

The nested enq operation remains simple: it appends the new node to the tail of the child queue 
(line \ref{inalg:cEnq}). The nested deq first locks the shared queue. Then, the next node to
return from deq is determined in lines~\ref{inalg:deqS} -- \ref{inalg:deqe}, as illustrated in Figure~\ref{fig:nested-queue}. As long as there are nodes in the shared queue that have not been dequeued, deq returns the value of the next such node but does not yet remove it from the queue (line~\ref{inalg:deqS}). Whenever the shared queue has been exploited, we proceed to traverse the parent transaction's local queue (line~\ref{inalg:deqP}), and upon exploiting it, perform the actual deq from the nested transaction's local queue (line~\ref{inalg:deqe}).
A commit appends (migrates) the entire local queue of the child  to the tail of the parent's local queue. The queue's validation always returns true: if it never invoked dequeue, its read set is empty, and otherwise, it had locked the queue.

We note that acquiring locks within nested transactions may result in deadlock. Consider the following scenario: Transaction $T_1$ dequeues from $Q_1$ and $T_2$ dequeues from $Q_2$, and then both of them initiate nested transactions that dequeue from the other queue ($T_2$ from $Q_1$ and vice versa). In this scenario, both child transactions will inevitably fail no matter how many times they are tried. To avoid this, we retry the child transaction only a bounded number of times, and if it exceeds this limit, the parent aborts as well and releases the locks acquired by it. Livelock at the parent level can be addressed using standard mechanisms (backoff, etc.).

To extend TDSL's skiplist with nesting we preserve its optimistic design. A child transaction maintains read- and write-sets of its own, and upon commit, merges them into its parent's sets. As in the queue, read operations of child transactions can read values written by the parent. Validation of the child's read-set verifies that the versions of the read objects have not changed.
The skiplist’s implementation is straightforward, and for completeness, we present its pseudo-code in the supplementary material.

\subsection{To Nest, or Not to Nest}\label{ssec:ubenchmark}
Nesting limits the scope of abort and thus reduces the overall abort rate. On the other hand, nesting introduces additional overhead. We now investigate this tradeoff.
\paragraph{Experiment setup} 
We run our experiments and measure throughput on an AWS m5.24xlarge instance with 2 sockets with 24 cores each, for a total of 48 physical cores. We disable hyperthreading.

We use a synthetic workload, where every thread runs 50,000 transactions, each consisting of 10 random operations on a shared skiplist followed by 2 random operations on a shared queue. Operations are chosen uniformly at random, and so are the keys for the skiplist operations. We examine three different nesting policies: (1) flat transactions (no nesting); (2) nesting skiplist operations and queue operations; and (3) nesting only queue operations. 

We examine two scenarios in terms of contention on the skiplist. In the low contention scenario, the skiplist's key range is from 0 to 50,000. In the second scenario, it is from 0 to 50, so there is high contention. Every experiment is repeated 10 times. 


\paragraph{Compared systems} 
We use the Synchrobench ~\cite{gramoli2015more}  framework in order to compare our TDSL to existing data structures optimized for running within transactions. Specifically, we run \estm~ (Elastic STM~\cite{elastic}) with the three transactional skiplists available as part of Synchronbench -- transactional friendly skiplist set, transational friendly optimized skiplist set, and transactional Pugh skiplist set --  and to the (single) available transactional queue therein. In all experiments we ran, the friendly optimized skiplist performed better than the other two, and so we present only the results of this data structure. This skiplist requires a dedicated maintenance thread in addition to the worker threads. To provide an upper bound on the performance of \estm, we allow it to use the same number of worker threads as TDSL plus an additional maintenance thread, e.g., we compare TDSL with eight threads to \estm~ with a total of nine. We note that \estm~ requires one maintenance thread per skip list; again, to favor \estm, we use a single skiplist in the benchmarks.

Synchrobench supports elastic transactions in addition to regular (opaque) ones, and also optionally supports multi-version concurrency control (MVCC) ~\cite{pstm1,pstm2}, which reduces abort rates on read-only transactions. We experiment with these two modes as well. 

We also ran our experiments on TL2 with the transactional friendly skiplist, but it was markedly slower than the alternatives, and in many experiments failed to commit transactions within the internally defined maximum number of attempts. We therefore omit these results.

\paragraph{Results}

\begin{figure*}[htb]
\begin{subfigure}[b]{0.3\textwidth}
\includegraphics[trim={0 0 0 0},clip,width=\columnwidth]{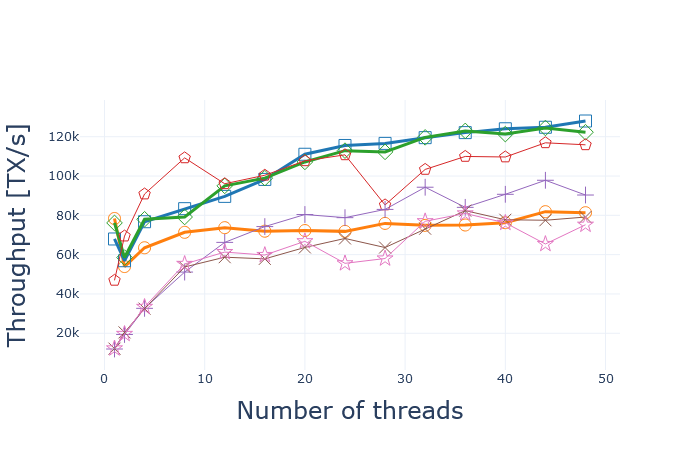}
\caption{Throughput, low contention} \label{sfig:50ktp}
\end{subfigure}
\begin{subfigure}[b]{0.3\textwidth}
\includegraphics[trim={0 0 0 0},clip,width=\columnwidth]{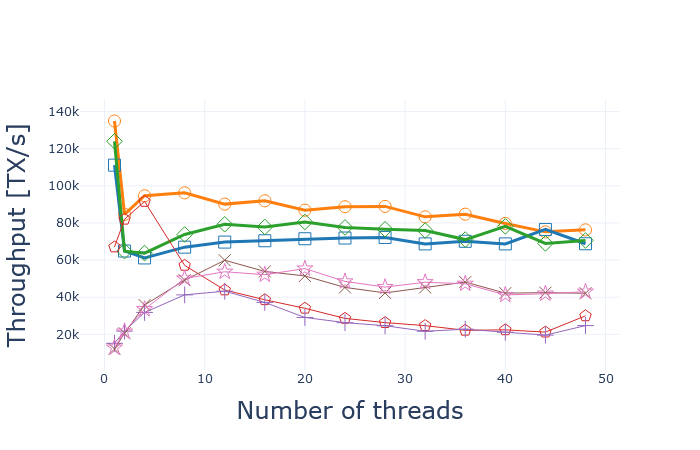}
\caption{Throughput, high contention}
\label{sfig:50tp}
\end{subfigure}
\begin{subfigure}[b]{0.3\textwidth}
\includegraphics[trim={0 0 0 0},clip,width=\columnwidth]{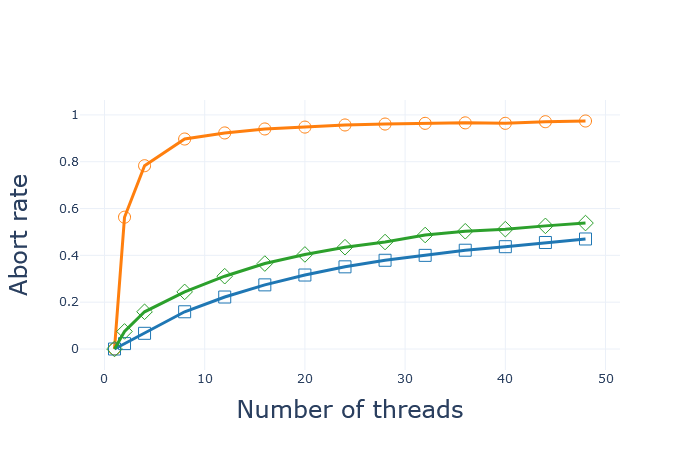}
\caption{Abort rate, low contention} \label{sfig:50kar}
\end{subfigure}
\begin{subfigure}[t]{1\textwidth} 
\includegraphics[width=\columnwidth]{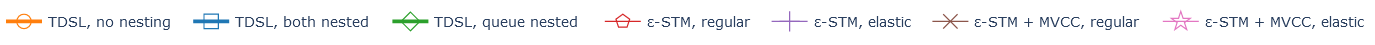}
\end{subfigure}
\caption{The impact of nesting in TDSL, compared to transactional friendly data structures on \estm.}
\label{fig:ubench}
\end{figure*}

Figure \ref{fig:ubench} shows the average throughput obtained.

In the low contention scenario (Figure \ref{sfig:50ktp}), nesting both queue and skiplist operations yields the best performance in the vast majority of data points. It improves throughput by 1.6x on average compared to flat transactions on 48 threads. It is worth noting that nesting provides the highest throughput without relaxing opacity like elastic transactions, and without keeping track of multiple versions of memory objects like MVCC. This is due to less work being wasted upon abort. Nesting queue operations seems to be the main reason for the performance gain compared to flat transactions, as nesting only queue operations yields comparable performance. In fact, nesting only the operations on the contended object may be preferable, as it provides the best of both worlds: low abort rates, as discussed later in this section, and less overhead around skiplist sub-transactions. The overhead difference is seen clearly when examining the performance of the two nesting variants of TDSL with a single thread, when nesting induces overhead and offers no benefits. 

We further investigate the effect of nesting via the abort rate, shown in Figure \ref{sfig:50kar} (for the low contention scenario). We see the dramatic impact of nesting on the abort rate. This sheds light on the throughput results. Nesting both skiplist and queue operations indeed minimizes the abort rate. However, the gap in abort rate does not directly translate to throughput, as it is offset by the increased overhead.  

In the high contention scenario (Figure \ref{sfig:50tp}), both DSs are highly contended, and nesting is harmful. The high contention prevents the throughput from scaling with the number of threads, and we observe degradation in performance starting from as little as 2 concurrent threads for TDSL, and between 4-12 concurrent threads for the other variants. From the abort rate point of view (graph omitted due to space limitations), the majority of transactions abort with as little as 4 threads regardless of nesting, and 80-90\% abort with 8 threads. Despite exhibiting the lowest abort rate, nesting all operations performs worse than other TDSL variants. In this scenario, too, nested TDSL performs better than the \estm~ variants spite being unfruitful compared to flat transactions.

%


Aborts on queue operations occur due to failures of \emph{nTryLock}, which has a good chance of succeeding if retried. On the other hand, aborts on nested skiplist operations are due to reading a higher version than the parent's VC. In such scenarios, the parent is likely to abort as well since multiple threads modify a narrow range of skiplist elements, hence an aborted child is not very likely to commit even if given another chance. 
Overall, we find that nesting the highly contended queue operations is more useful than nesting map operations -- even when contended. Thus, contention alone is not a sufficient predictor for the utility of nesting. Rather, the key is the likelihood of the failed operation to succeed if retried.


%% file: NIDS.tex
\section{NIDS Case Study}\label{NIDS}

We conduct a case study of parallelizing a full-fledged network intrusion detection system using memory transactions. In this section we provide essential background for multi-threaded IDS systems, describe our NIDS software and point out candidates for nesting.


Intrusion detection is a basic security feature in modern networks, implemented by popular systems such as Snort 
\cite{Snort}, Suricata \cite{Suricata}, and Zeek \cite{Bro}. As network speeds increase and bandwidth grows,  NIDS performance becomes paramount, and multi-threading becomes instrumental~\cite{Intruder}.

\begin{figure}[h]
\centering
\includegraphics[width=0.8\columnwidth]{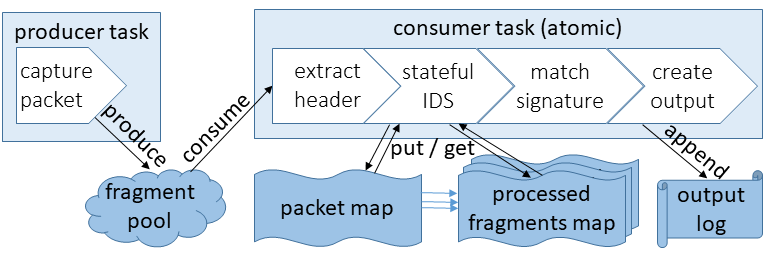}
\caption{Our NIDS benchmark: tasks and data structures.}
\label{fig:NIDS}
\end{figure}

\textbf{Multi-threaded NIDS.}
We develop a multi-threaded NIDS benchmark. The processing steps executed by the benchmark follow the 
description in~\cite{Intruder}.
As illustrated in Figure~\ref{fig:NIDS}, our design 
employs two types of threads. First, \emph{producers} simulate the \emph{packet capture} 
process of reading packet fragments off a network interface. In our benchmark, we do not use an actual network,
and so the producers generate the packets and push MTU-size packet fragments into a shared producer-consumer pool called the \emph{fragments pool}. 
The rationale for using dedicated threads for packet capture is that -- in a real system -- the amount of work these threads have scales with network resources rather than compute and DRAM resources. 
In our implementation, the producers simply drive the benchmark and do not do any actual work. 

Packet processing is done exclusively by the \emph{consumer} threads, each of which consumes 
and processes a single packet fragment from the shared pool.  Algorithm \ref{alg:nestSuggestion} describes  the consumer's code.
To ensure consistency, each consumer executes as a single atomic transaction. 
It begins by performing \emph{header extraction}, namely, extracting information from the link layer header.
The next step is called  \emph{stateful IDS}; 
it consists of packet reassembly and detecting violations of protocol rules.
Reassembly uses a shared \emph{packet map} associating each packet with its own  
shared \emph{processed fragment map}.
The first thread to process a fragment pertaining to a particular packet creates the packet's fragment map whereas other threads append fragments to it. Similarly, 
only the thread that processes a packet's last fragment continues to process the packet, 
while the remaining threads move on to process other fragments from the pool. 
By using atomic transactions, we guarantee that indeed there are unique ``first'' and ``last'' threads
and so consistency is preserved.  

\begin{algorithm}
\caption{Consumer code}
\begin{algorithmic}[1]\small
\State $\var{f} \gets$ fragmentPool.consume()
\State process headers of $\var{f}$
\State fragmentMap $\gets$ packetMap.get($\var{f}$) \label{inalg:atBegin} \Comment Start nested TX 
\If    {fragmentMap  $=  \bot$ }
\State		fragmentMap $\gets$ new map
\State		packetMap.put($\var{f}$, fragmentMap) \Comment End nested TX 
\EndIf \label{inalg:atEnd} 
\State fragmentMap.put($\var{f.id}$, $\var{f}$)
\If {$\var{f}$ is the last fragment in packet}
\State reassemble and inspect packet \Comment Long computation
\State log the result						\Comment Nested TX
\EndIf
\end{algorithmic}
\label{alg:nestSuggestion}
\end{algorithm}

The thread that puts together the packet proceeds to the \emph{signature matching} phase, whence the reassembled packet's content is tested against a set of logical predicates; if all are satisfied, the signature matches.
This is the most computationally expensive stage~\cite{Intruder}. 
Finally, the thread generates a packet trace and writes it to a shared log. 

As an aside, we note that our benchmark performs five of the six processing steps detailed in~\cite{Intruder};  
the only  step we skip is 
\emph{content normalization}, which unifies the representations of packets that use different application-layer protocols. This phase is redundant in our solution since we use a unified packet representation to begin with.
In contrast, the \emph{intruder} benchmark in STAMP~\cite{STAMP} implements a more limited functionality,
consisting of packet reassembly and na\"ive signature matching: 
threads obtain fragments from their local states (rather than a shared pool),
signature matching is lightweight, and no packet traces are logged. 
This results in significantly shorter transactions than in our solution. 

\textbf{Nesting.} 
We identify two candidates for nesting. 
The first is the logging operation given that logs are prone to be highly contended. Because in this application the logs are write-only, transactions abort only when they contend to write at the tail and not because of consistency issues. Therefore, retrying the nested transaction amounts to 
retrying to acquire a lock on the tail, which is much more efficient than restarting the transaction. 

Second, when a packet consists of multiple fragments, its entry in the packet map is contended. In particular, for every fragment, a transaction checks whether an entry for its packet exists in the map, and creates it if it is absent. Nesting lines \ref{inalg:atBegin} - \ref{inalg:atEnd} of Algorithm \ref{alg:nestSuggestion} may thus prevent aborts. 

%% file: additionalDS.tex
\section{Additional Nestable Data Structures}\label{MoreDS}
Transactions may span multiple objects of different types. Every DS implements the methods defined by its type (e.g., deq for queue), as well as methods for validation, migrating a child transaction's state to its parent, and committing changes to shared memory. We extend 
our Java TDSL with three widely used data structures -- a producer-consumer pool, a log, and a stack (found in the full version of this paper due to space limitation). For each, we first describe the transactional implementation and then how nesting is achieved.


\paragraph{Producer-consumer Pool} Like many other applications, our NIDS benchmark uses a producer-consumer pool. Such pools are also a cornerstone in architectures like SEDA \cite{SEDA}. They scale better than queues because they don't guarantee order~\cite{gidron2012salsa,basin2011cafe}. We support a bounded-size transactional producer-consumer pool  consisting of a pre-defined number of \textit{slots, K}. The \emph{produce} operation finds a free slot and inserts a consumable object into it, while the \emph{consume} operation finds a produced object and consumes it. 
Our pool guarantees that if there is an available slot for consumption and any number of consumer threads, then at least one of the consumers will consume it, and similarly for free slots and producers. 

Algorithm \ref{alg:ppc} presents our nested operations for the producer-consumer pool. The parent operations are very similar and are deferred to the supplementary material. 
We assume that the consumer functions passed to the consume method do not have any side effects that 
are visible in shared state before the transaction commits.   

Similarly to  deq,  consume also warrants pessimistic concurrency control, as each object can be consumed at most once. But the granularity of locking is finer-grain, namely, consume locks a single slot rather than the entire pool, which allows much more parallelism. Produce is also pessimistic at the same granularity, ensuring that the same slot is not concurrently used by multiple threads.
More specifically, we assign a state to each slot in the pool, as follows: $\bot$ means that the slot is free. A slot is in the \emph{locked} state if there is an ongoing transaction that uses it. A \emph{ready} slot is available to be consumed. We implement the  methods 
\emph{getFreeSlot} and \emph{getReadySlot}, which atomically find and lock a free or ready slot, respectively, using CAS. The \emph{changeState} method executes a state transition atomically using a CAS. 
We use these methods to ensure that a ready slot is populated by at most one transaction and a produced item is consumed at most once: 
We keep track of slots that are locked by the current transaction in two sets: \textit{produced} and \textit{consumed}. A slot's state changes from \textit{locked} to \textit{ready} upon successful commit of a parent locking transaction, and changes from \textit{locked} to $\bot$ either upon successful commit or upon cancellation 
as we describe next. Upon abort, every slot's state reverts to its previous state.

We now explain how we use cancellation for liveness. Consider a pool of size K and a transaction $\var{T_1}$ that performs $K+1$ produce operations, each followed by a consume operation. If every operation locks a slot until the commit time, such a transaction cannot proceed past the K'th consume, despite the fact that the transaction respects the semantics of the data structure. To mitigate this effect, our implementation consumes slots that were produced within the same transaction before locking additional \textit{ready} slots, and releases the lock of any consumed slot by setting its state to $\bot$. This way, consumed slots cancel out with produced slots within the same transaction. Since cancellation occurs in thread-local state, which is not accessed by more than one thread,  correctness is trivially preserved. 

\begin{algorithm}[h]
\caption{Nested operations for producer-consumer pool}
\label{alg:ppc}
\begin{algorithmic}[1]\footnotesize
\State \textbf{PCPool} P
\Indent 
	\State \textbf{Shared:} Slots[K]
	\State \textbf{Thread local:} parentProduced, parentConsumed,
	\Indent
	\State childProduced, childConsumed, childConsumedParent
	\EndIndent
\EndIndent

\Procedure{nConsume}{consumer}
	\If{childProduced.size$>$0}\label{inalg:ccbegin1}
		\State n $\gets$ childProduced.pop()
		\State consumer.consume(n.val)
		\State n.changeState($\bot$)\label{inalg:ccend1} \Comment{Cancellation}
	\ElsIf{parentProduced.size>0}\label{inalg:childConsumeFromParentBegin}
		\State n $\gets$ parentProduced.getNext()
		\State consumer.consume(n.val)
		\State childConsumedParent.add(n)\label{inalg:childConsumeFromParentEnd}	
	\Else
		\State  n $\gets$ \Call{P.getReadySlot}{\null} 
																								\label{inalg:childConsumeFromSharedBegin}
		\State consumer.consume(n.val)
		\State childConsumed.add(n)\label{inalg:childConsumeFromSharedEnd}
	\EndIf
\EndProcedure
\Procedure{nProduce}{val} 
	\State n $\gets$ \Call{P.getFreeSlot}{\null} \Comment{Changes state to \emph{locked}}
	\State n.val $\gets$ val
	\State childProduced.add(n)
\EndProcedure

\Procedure{migrate}{} \Comment On child commit
	\State merge childConsumed into parentConsumed
	\State merge childProduced into parentProduced
	\ForAll{n in childConsumedParent} \label{l:childConsumedParentStart}
		\State remove n from parentProduced
		\State n.state $\gets \bot$ 
	\EndFor \label{l:childConsumedParentEnd}
\EndProcedure

\Procedure{validate}{} \Comment Always succeeds
\State \Return true
\EndProcedure
\end{algorithmic}
\end{algorithm}

As in other nestable data structures, the child's local state is structured like the parent's. 
When nesting, the cancellation logic is expanded. The consume operation first tries to consume from child-local produced slots (lines \ref{inalg:ccbegin1}--\ref{inalg:ccend1}), then from parent-local ones (lines \ref{inalg:childConsumeFromParentBegin}--\ref{inalg:childConsumeFromParentEnd}) and only then locks a slot (line \ref{inalg:childConsumeFromSharedBegin}). We keep track of slots that were produced by the parent and consumed by the child in \emph{childConsumedFromParent}. 
The state of such slots changes back to $\bot$ when the child commits (lines~\ref{l:childConsumedParentStart}--\ref{l:childConsumedParentEnd}).
 Additionally, at the end of the child transaction, the \textit{produced} and the \textit{consumed} sets of the child  are merged with the parent's sets.
Because access to slots is pessimistic, our pool involves no speculative execution, and so validate always returns true.


\paragraph{Log} Logs are commonly used for record-keeping of events in a system, as occurs in our NIDS benchmark. Recently, they are also popular for ledger transaction ordering. Logs are unique because their prefixes are immutable, whereas their tail is an ever-changing contention point among concurrent write operations.
A log has 2 operations: \emph{read(\emph{i})} and \emph{append(val)}. Read($\var{i}$) returns the value in position $\var{i}$ in the log or $\bot$ if it has not been created yet. Append($\var{val}$) appends $\var{val}$ to the log. 

Log reads never modify the state of the log, which lends itself to an optimistic implementation. Append, on the other hand, is more amendable to a pessimistic solution, since the semantics of the log imply that only one of any set of interleaving appending transactions may successfully commit.

Read-only transactions that do not reach the end of the log are not subject to aborts. A transaction that either reaches the end (i.e., read($\var{i}$) returns $\bot$) or appends is prone to abort. The log’s implementation uses local logs (at both a parent and a child transaction) to keep track of values appended during the transaction; each has a Boolean variable indicating whether it encountered an attempt to read beyond the end of the shared log. 
The local log also records the length of the log at the time of the first access to the log. 
A read($\var{i}$) operation reads through the shared and the parent logs, and in the presence of nesting, the child log as well. Append($\var{val}$) locks the log and appends $\var{val}$ to the current transaction's local log. For completeness, we provide the log’s pseudocode in the supplementary material. 

The correctness of our log stems from the following observations: first, two writes cannot interleave, since a write is performed only if a lock had been acquired. Second, a transaction will commit if it either had not read or modified (by appending to) the end of the log, or if there hadn't been later writes. Finally, since the log is modified at commit time, opacity is preserved, i.e., no transaction sees inconsistent partial updates.

%% file: Evaluation.tex
\section{NIDS Evaluation}\label{Evaluation}
We now experiment with nesting in the NIDS benchmark. We detail our evaluation methodology in Section \ref{sub:setup} and present quantitative results in Section \ref{sub:results}.

\subsection{Experiment Setup}
\label{sub:setup}

\begin{figure*}[htb]
\begin{subfigure}[b]{0.3\textwidth}
\includegraphics[trim={0 0 0 0},clip,width=\columnwidth]{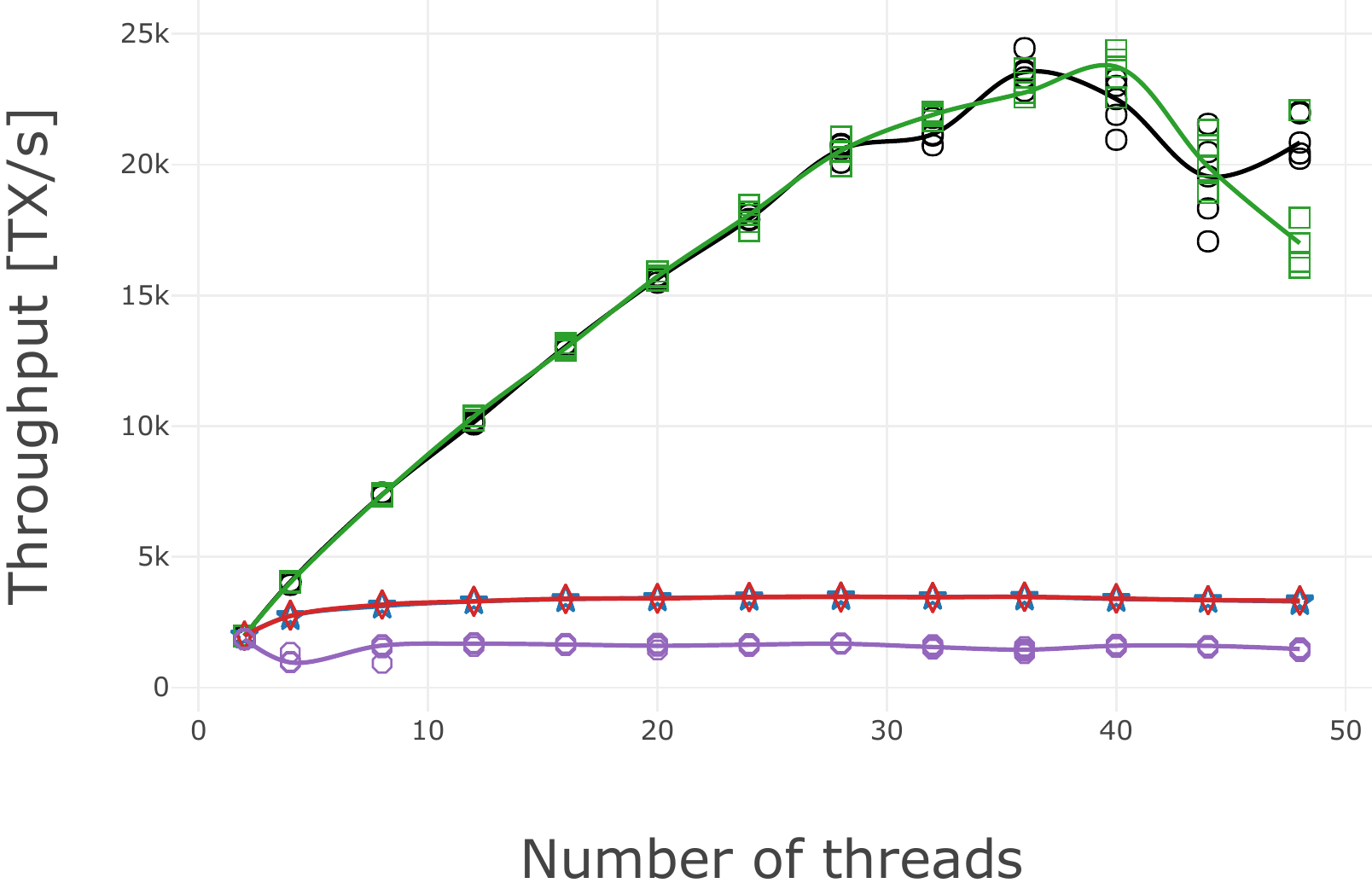}
\caption{Throughput, one fragment per packet} \label{fig:aws_1_1_TP}
\end{subfigure}
\begin{subfigure}[b]{0.3\textwidth}
\includegraphics[trim={0 0 0 0},clip,width=\columnwidth]{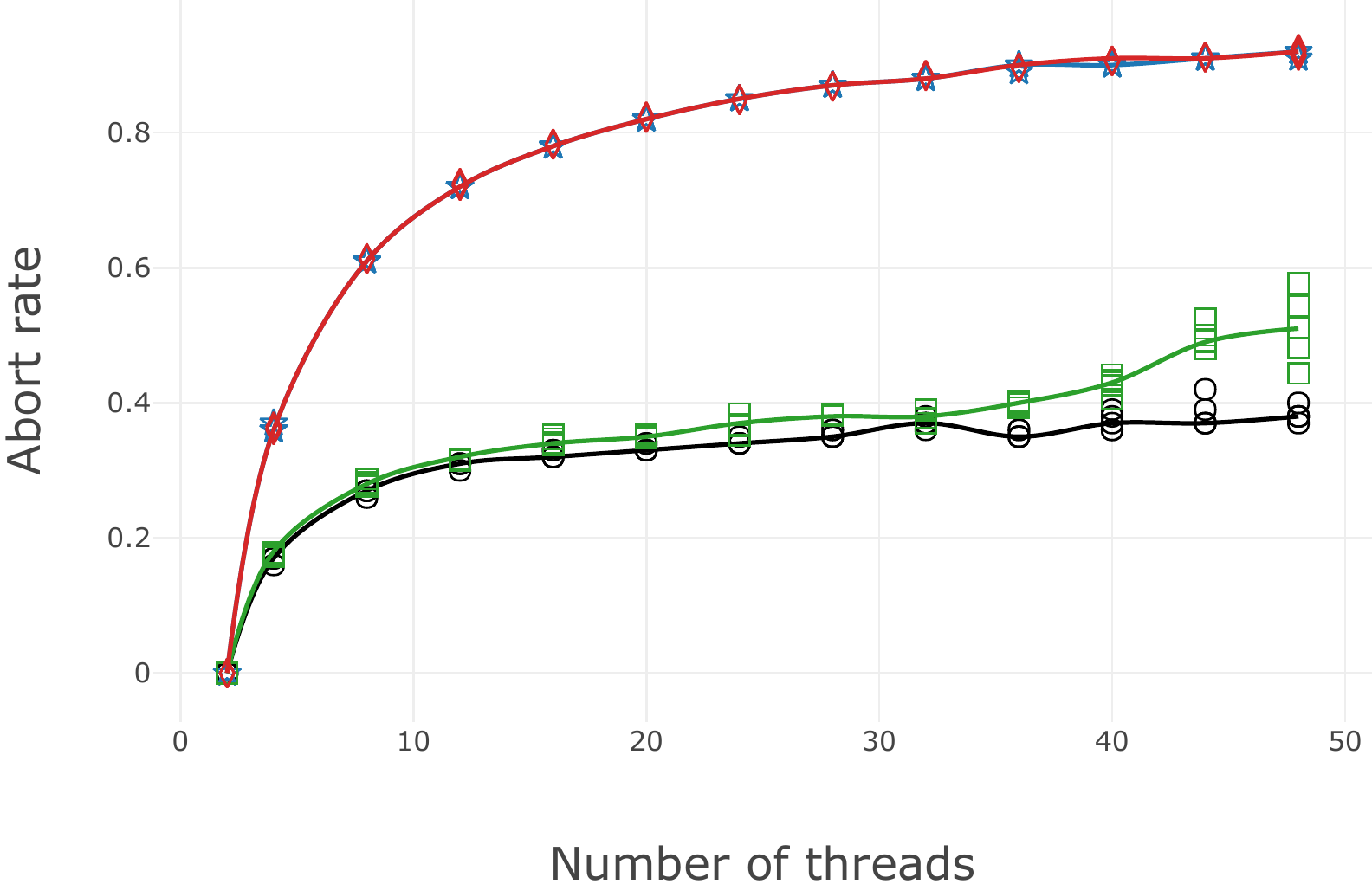}
\caption{Abort rate, one fragment per packet} \label{fig:aws_1_1_AR}
\end{subfigure}
\begin{subfigure}[b]{0.3\textwidth}
\includegraphics[trim={0 0 0 0},clip,width=\columnwidth]{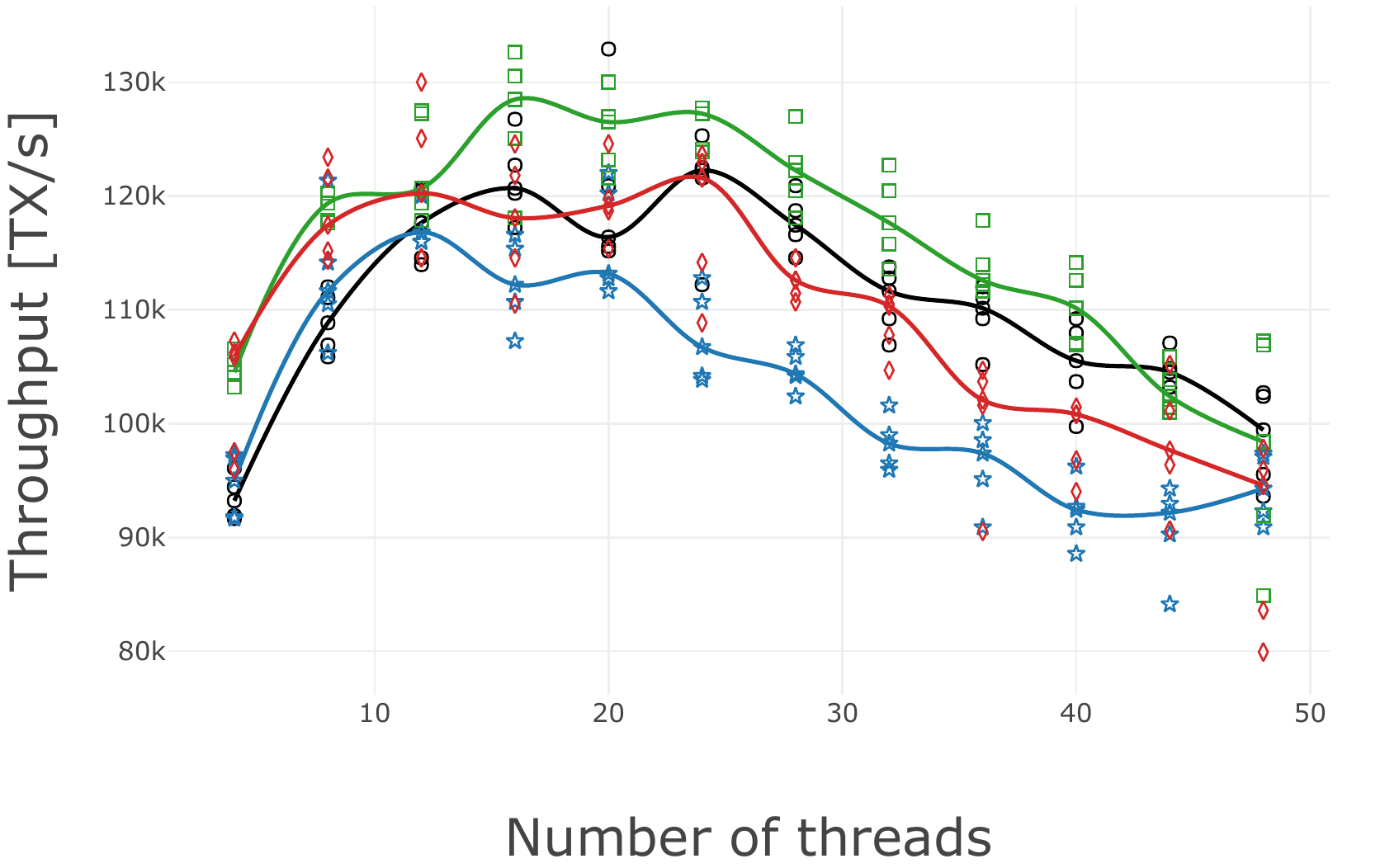}
\caption{Throughput, 8 fragments per packet} \label{fig:noTL2_TP}
\end{subfigure}
\begin{subfigure}[t]{1\textwidth}
\includegraphics[width=\columnwidth]{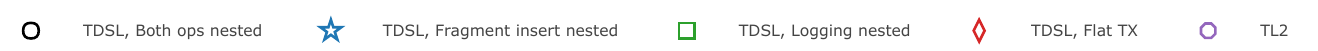}
\end{subfigure}
\caption{NIDS experiments results.}
\end{figure*}


Our baseline is TDSL without nesting, which is the starting point of this research. 
We also compare to the open source Java STM implementation of TL2 by Korland et al.~\cite{deuceTL2}, as well as \estm~\cite{elastic} and \vcc~\cite{pstm1, pstm2}. The results we obtained for \estm~ and \vcc~ were very similar to those of TL2 and are omitted from the figures to avoid clutter. Note that the open-source implementations of \estm~ and PSTM optimize only data structures that contain integers; they use bare-STM implementations for data structures holding general objects, as the data structures in our benchmark do. This explains why their performance is sub-optimal in this benchmark.

We experiment with nesting each of the candidates identified in Section \ref{NIDS} (put-if-absent to the packetMap and updating the log), and also with nesting both. Our baseline executes \emph{flat transactions}, i.e., with no nesting. In TDSL, the packet pool is a producer-consumer pool, the map of processed packets is a skiplist of skiplists, and the output block is a set of logs. For TL2, the packet pool is implemented with a fixed-size queue, the packet map is an RB-tree of RB-trees, and the output log is a set of vectors. We use the implementations provided in \cite{JSTAMP} without modification.

The experiment environment is the same as for the microbenchmark described in Section \ref{ssec:ubenchmark}. We repeated the experiment on an in-house 32-core Xeon machine and observed similar trends; these results are omitted.
We run each experiment 5 times and plot all data points, connecting the median values with a curve.

We conduct two experiments. In the first, each packet consists of a single fragment, there is one producer thread, and we scale the number of consumers. In the second experiment, there are 8 fragments per packet and as we scale the number of threads, we designate half the threads as producers. We experimented also with different ratios of producers to consumers, but this  did not seem to have a significant effect on performance or abort rates, 
so we stick to one configuration in each experiment. The number of fragments per packet governs  contention: If there are fewer fragments then more threads try to write to logs simultaneously.  With more fragments, on the other hand, there are more put-if-absent attempts to create maps. 

\subsection{Results}\label{sub:results}

\begin{figure}[h]
\centering
\includegraphics[trim={0 0 0 0},clip,width=0.65\columnwidth]{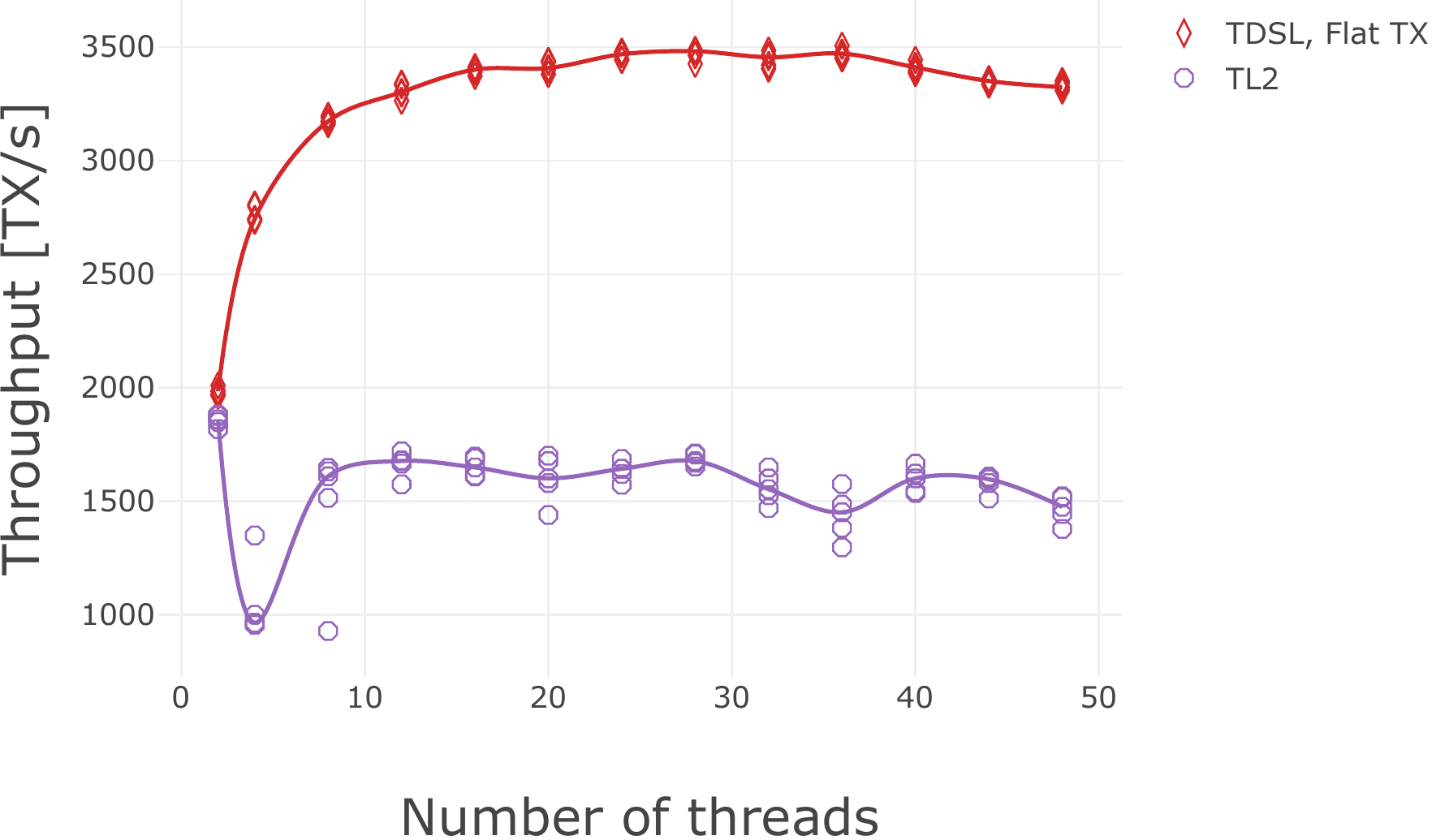}
\caption{Throughput of TL2 and flat transactions in TDSL, a single producer and one fragment per packet.}
\label{fig:onlyflat}
\end{figure}

\textbf{Performance.} Figures \ref{fig:aws_1_1_TP} and  \ref{fig:aws_1_1_AR} show the throughput and abort rate in a run with 1 fragment per packet and a single producer. Whereas the performance of all solutions is similar when we run a single  consumer, performance differences become apparent as the number of threads increases. For flat transactions (red diamonds), TDSL's throughput is consistently double that of TL2 (purple octagons), as can be observed in Figure \ref{fig:onlyflat}, which zooms in on these two curves in the same experiment. 
We note that the TDSL work \cite{TDSL} reported better performance improvements over TL2, but they ran shorter transactions that did not write to a contended log at the end, where TDSL's abort rate remained low. In contrast, our benchmark's long transactions  result in high abort rates in the absence of nesting.
Nesting the log writes (green squares) improves throughput by an additional factor of up to 6, which is in 
line with the improvement of TDSL over TL2 reported in \cite{TDSL}, 
and also reduces the abort rate by a factor of 2. 
The packet map is not contended in this experiment, and so transactions with nested insertion to the map behave similarly to flat ones (in terms of both throughput and abort rate).


Figure \ref{fig:noTL2_TP} shows the results in experiments with 8 fragments per packet. For clarity, we omit TL2 from this graph because it performs 6 times worse than the lowest alternative. 
Here, too, the best approach is to nest only log updates, but the impact of such nesting is less significant in this scenario, improving throughput only by about 20\%. This is because with one fragment per packet, every transaction tries to write to the log,  whereas with 8, only the last fragment’s transaction does, reducing contention on the log. Nevertheless, the effect of nesting log updates is more significant as it reduces the number of aborts by a factor of 3, and thus saves work.

Unlike in the 1-thread scenario, with 8 threads, there is contention on the put-if-absent to the fragment map, and so nesting this operation reduces aborts. At first, it might be surprising that flat transactions perform better than ones that nest the put-if-absent despite their higher abort rate. 
However, the abort reduction has a fairly low impact since this operation is performed early in the transaction. Thus, the overhead induced by nesting exceeds the benefit of not repeating the earlier part of the computation. The effect of this overhead is demonstrated in the difference in performance between nesting both candidates (black circles) and nesting only the log writes (green squares).

\begin{table*}[hbt]
\centering
\begin{tabular}{|c|c|c|c|c|c|}
\hline
 & TL2 & \begin{tabular}[c]{@{}c@{}}TDSL flat\end{tabular} & \begin{tabular}[c]{@{}c@{}}TDSL nesting log\end{tabular} & TDSL nesting put-if-absent & TDSL nesting both \\ \hline
1 fragment & 1.6K / 8 & 3.5K / 28 & 23.5K /40  & 3.5K / 28 & 23.5K / 36 \\ \hline
8 fragments & 24K / 4& 122K / 24 & 127K / 24 & 113K / 20 & 122K / 24 \\ \hline
\end{tabular}
\caption{Scalability: peak performance (tx/sec) / number of threads where it is achieved.}
\label{tab:scaling}
\end{table*}

%
%
\textbf{Scaling.} 
Not only does nesting have a positive effect on performance, it improves scalability as well. For instance, Figure \ref{fig:aws_1_1_TP} shows that throughput increases linearly all the way up to 40 threads when nesting the logging operation, whereas flat nesting, as can be seen in Figure \ref{fig:onlyflat}, peaks at 28 threads but saturates already at 16.
Table \ref{tab:scaling} summarizes the scaling factor in both experiments. 




%% file: related_work.tex
\section{Related Work}\label{related}
\textbf{Transactional data structures.}
Since the introduction of TDSL~\cite{TDSL} and
STO~\cite{itiswhatitis}, transactional libraries got a fair bit of
attention~\cite{FL-wait-free1, loft,FL-lock-free1,
defer}.
Other works have focused on wait-free
\cite{FL-wait-free1} and lock-free
\cite{loft,FL-lock-free1} implementations (as opposed
to TDSL and STO's lock-based approach).
Such algorithms are interesting from a theoretical point of
view, but provide very little performance benefits, and
in some cases can even yield worse results than lock-based
solutions~\cite{ennals2006software, dice2007understanding}.


\remove{Huang et al.~\cite{OCC} tailor STO's version management for
in-memory databases; our evaluation focuses on other use cases.}
Lebanoff et al.~\cite{defer} introduce a trade-off between low abort
rate and high computational overhead.
By restricting their attention to static transactions, they are able
to perform scheduling analyses in order to reduce the overall system
abort rate.
We, in contrast, support dynamic transactions.

Transactional boosting and its follow-ups~\cite{optimisticTXBoosting, herlihy2008transactional} offer generic approaches for making
concurrent data structures transactional.
However, they do not exploit the structure of the transformed data
structure, and instead rely on semantic abstractions like
compensating actions and abstract locks. 

Some full-fledged STMs incorporate optimization for specific data structures. For instance, \estm~ \cite{elastic} and \vcc~ \cite{pstm1, pstm2} support elastic transactions on search data structures. Note, however, that unlike closed nesting, elastic transactions relax transactional semantics. PSTM allows programmers to select the concurrency control mechanism (MVCC or single-version) and the required semantics (elastic or regular) for each transaction. While this offers a potential for performance gains, our results in Section \ref{ssec:ubenchmark} have shown that nesting outperforms all of the approaches. 

PSTM improves on SwissTM \cite{dragojevic2009stretching}, which has featured other optimizations in order to support longer transactions than implementations that preceded it, like a contention manager and mixed concurrency control, and showed 2-3x better performance compared to TL2 \cite{TL2} and TinySTM \cite{felber2008tinystm} and good scalability up to 8 threads. These optimizations are orthogonal to nesting. 


In this paper we extend the original TDSL
algorithm~\cite{TDSL}.
To the best of our knowledge, none of the previous works on transactional data structures provide nesting. 

\textbf{Chopping and nesting.} 
Recent works introduced the concept of
\emph{chopping}~\cite{db1,db2,db3}, which splits up transactions
in order to reduce abort rates. Chopping and the similar concept of elastic transactions~\cite{elastic} were recently adopted in
transactional memory~\cite{DRP,htmChop,defer}.
The high-level idea of chopping is to divide a transaction into
a sequence of smaller ones and commit them one at a time.
While atomicity is eventually satisfied (provided that all
transactions eventually commit), this approach forgoes isolation and
consistency, which nesting preserves.

While some previous work on supporting nesting in generic STMs was
done in the past~\cite{openNestingUse1, turcu2012closed,
MOSSHOSKING, nesting1}, we are not aware of any previous work implementing \emph{closed nesting} in a non-distributed sequential STM. This might be due to the fact that the benefit of closed nesting is in allowing longer transactions whereas STMs are not particularly suitable for long transactions in any case, and the extra overhead associated with nesting might be excessive when read- and write-sets are large as in general purpose STMs. Our solution is also the first to introduce nesting into transactional data structure libraries, and thus the first to exploit the specific structure and semantics of data structures for efficient nesting implementations. Because our data-structures use diverse concurrency control approaches, we had to develop nesting support for each of them. An STM using any of these approaches (e.g., fine-grain commit-time locking with read-/write-sets) can mimic our relevant technique (e.g., closed-nesting can be supported in TL2 using a similar scheme to the one we use in maps). 

  

%% file: Conclusion.tex
\section{Conclusion}

\label{sec:conclusion}


The TDSL approach enables high-performance software transactions by restricting transactional access to a well-defined set of data structure operations. Yet in order to be usable in practice, a TDSL needs to be able to sustain long transactions, and to offer a variety of data structures. In this work, we took a step towards boosting the performance and usability of TDSLs, allowing them to support complex applications. A key enabler for long transactions is nesting, which limits the scope of aborts without changing the semantics of the original transaction.

We have implemented a Java TDSL with built-in support for nesting in a number of data structures. We conducted a case study of a complex network intrusion detection system running long transactions. We found that nesting improves performance by up to 8x,  and the nested TDSL approach outperforms the general-purpose general-purpose STM by up to 16x. We plan to make our code (both the library and the benchmark) available in open-source.

%% file: DISCAppendix.tex
\section{Supplementary Material: Additional Pseudocode}\label{appx}

\begin{algorithm}[h]
\begin{algorithmic}[1]
\State \textbf{Skiplist}
\Indent 
	\State \textbf{Shared}
	\Indent	
		\State sharedSkiplist 
	\EndIndent
	\State \textbf{Thread local}
	\Indent
		\State parentReadSet, parentWriteSet,
		\State childReadSet, childWriteSet 
	\EndIndent
\EndIndent

\Procedure {nGet}{key}
	\State add key to childReadSet
    \If{key $\in$ childWriteSet}
    	\State \Return value from childWriteSet
    \ElsIf{key $\in$ parentWriteSet}
    	\State \Return value from parentWriteSet
    \ElsIf{key $\in$ sharedSkiplist}
    	\State \Return value from sharedSkiplist
    \EndIf
    \State \Return $\bot$
\EndProcedure
\Procedure {nPut}{key,value}
	\State add key to childWriteSet
\EndProcedure

\Procedure {nRemove}{key}
	\State add $<$ key, remove $>$ to childWriteSet
\EndProcedure

\Procedure {validate}{}
	\ForAll{obj in childReadSet}
		\If{obj.version $>$ parent version}
			\State  \textbf{abort}
		\EndIf
	\EndFor
\EndProcedure
\Procedure {migrate}{}
	\State merge childWriteSet into parentWriteSet
	\State merge childReadSet into parentReadSet
\EndProcedure
\end{algorithmic}
\caption{Nested operations on skiplists}
\label{alg:list}
\end{algorithm}

\begin{algorithm}[h]
\caption{Nestable transactional log}
\label{alg:log}
\begin{algorithmic}[1]
\State \textbf{Log}
\Indent
	\State \textbf{Shared}
	\Indent	
		\State sharedLog 
	\EndIndent
	\State \textbf{Thread local}
	\Indent
	 \State parentLog, childLog,
 \State readAfterEnd (initially \textit{false}),
 \State initLen (initial size of sharedLog)
	\EndIndent
\EndIndent

\Procedure{append}{val}
	\State \hspace{0.5cm}\Comment Parent code
	\State \Call{tryLock}{\null}
	\State \Call{parentLog.append}{val}
\EndProcedure

\Procedure{read}{$i$}
	\State \hspace{0.5cm}\Comment Parent code 
	\If	{$i \in $ sharedLog}
	\State \Return  sharedLog[$i$]
	\Else
		\State readAfterEnd $\gets$ true
		\If{$i \in$ parentLog}
		 \State \Return  parentLog[$i$]
		 \Else
		 	\State \Return $\bot$
		 \EndIf
	\EndIf
\EndProcedure

\Procedure{nAppend}{val}
	\State \hspace{0.5cm}\Comment Nested (child) code 
	\State \Call{nTryLock}{\null}
	\State \Call{childLog.append}{val}
\EndProcedure
\Procedure{nRead}{$i$}
	\State \hspace{0.5cm}\Comment Nested (child) code 
	\If	{$i \in $ sharedLog}
	\State \Return  sharedLog[$i$]
	\Else
		\State readAfterEnd $\gets$ true 
		\If{$i \in$ parentLog}
		 \State \Return  parentLog[$i$]
		\ElsIf {$i \in$ childLog}
			\State \Return childLog[$i$]
			\Else
				\State \Return $\bot$
		\EndIf
	\EndIf
\EndProcedure

\Procedure{migrate}{}
	\State \hspace{0.5cm} \Comment Occurs on child commit
	\State append childLog to parentLog
\EndProcedure

\Procedure{validate}{}
\If{readAfterEnd $\; \land$ sharedLog exceeds initLen}
\State \Return abort
\EndIf
\State \Return true
\EndProcedure

\end{algorithmic}
\end{algorithm}

\begin{algorithm}[h]
\caption{Parent operations for producer-consumer pool}
\label{alg:ppc_Parent}
\begin{algorithmic}[1]
\State \textbf{Producer-consumer Pool}
\Indent 
	\State \textbf{Shared}
	\Indent	
		\State Slots[K] 
	\EndIndent
	\State \textbf{Thread local}
	\Indent
 \State parentProduced, parentConsumed 
 	\EndIndent
\EndIndent

\Procedure{Produce}{val} 
	\State n $\gets$ \Call{getFreeSlot}{\null}
	\State \hspace{0.5cm} \Comment{n.state $\gets$ \emph{locked}}
	\State n.val $\gets$ val
	\State parentProduced.add(n)
\EndProcedure
\Procedure{Consume}{consumer}
	\If{parentProduced.size $>$ 0}
		\State n $\gets$ parentProduced.pop()
		\State consumer.consume(n.val)
		\State n.changeState($\bot$) \label{inalg:cancel}\Comment{Cancellation}
	\Else
		\State  n $\gets$ \Call{getReadySlot}{\null}
		\State \hspace{0.5cm}\Comment{n.state $\gets$ \emph{locked}}
		\State consumer.consume(n.val)
		\State parentConsumed.add(n)
	\EndIf
\EndProcedure
\end{algorithmic}
\end{algorithm}